\documentclass[runningheads]{llncs}
\usepackage[T1]{fontenc}
\usepackage{graphicx}
\usepackage{booktabs}
\usepackage{amsmath}
\usepackage[misc]{ifsym}
\usepackage{hyperref}       
\usepackage{url}            
\usepackage{subcaption}
\usepackage{multirow}
\usepackage{booktabs}
\usepackage{float}
\usepackage[normalem]{ulem}
\usepackage{enumitem}

\hypersetup{
   colorlinks,
   menucolor=black,
   linkcolor=black,
   citecolor=red,
   urlcolor=blue
}

\begin{document}

\title{Membership Inference Attacks Beyond Overfitting}

\titlerunning{Membership Inference Attacks Beyond Overfitting}

\author{Mona~Khalil\inst{1} 
\and Alberto Blanco-Justicia\inst{1} \and Najeeb Jebreel\inst{1} \and Josep Domingo-Ferrer\inst{1,2}}

\authorrunning{Mona Khalil {\em et al.}}

\institute{Universitat Rovira i Virgili,\\ Department of Computer Engineering and Mathematics,\\
        CYBERCAT-Center for Cybersecurity Research of Catalonia,\\ Av. Pa\"{\i}sos Catalans 26, 43007 Tarragona, Catalonia\\
\email{\{mona.khalil, alberto.blanco, najeeb.jebreel, josep.domingo\}@urv.cat} \and LAAS-CNRS, Universit\'e de Toulouse, 7 Av. du Colonel Roche, 31400 Toulouse, France}

\maketitle              

\begin{abstract}
Membership inference attacks (MIAs) against machine learning (ML) models aim to determine whether a given data point was part of the model training data.
These attacks may pose significant privacy risks to individuals whose \textit{sensitive} data were used for training, which motivates the use of defenses such as differential privacy, often at the cost of high accuracy losses.
MIAs exploit the differences in the behavior of a model when making predictions on samples it has seen during training (\textit{members}) versus those it has not seen (\textit{non-members}).
Several studies have pointed out that model overfitting is the major factor contributing to these differences in behavior and, consequently, to the success of MIAs.
However, the literature also shows that even non-overfitted ML models can leak information about a small subset of their training data.
In this paper, we investigate the root causes of membership inference vulnerabilities beyond traditional overfitting concerns and suggest targeted defenses.
We empirically analyze the characteristics of the training data samples vulnerable to MIAs in models that are not overfitted (and hence able to generalize). 
Our findings reveal that these samples are often outliers within their classes ({\em e.g.}, noisy or hard to classify).
We then propose potential defensive strategies to protect these vulnerable samples and enhance the privacy-preserving capabilities of ML models. 
Our code is available at \url{https://github.com/najeebjebreel/mia_analysis}.
\end{abstract}

\section{Introduction}
\label{sec:introduction}
Machine learning (ML) has demonstrated remarkable performance across a wide range of tasks~\cite{he2016deep,devlin2018bert,miotto2018deep}.
This success is mainly attributed to the availability of large and diverse data for training, along with advances in learning algorithms and computational capabilities. 

However, training data often contain sensitive information related to individuals, such as personal photos~\cite{kemelmacher2016megaface}, confidential texts~\cite{carlini2021extracting}, clinical records~\cite{kourou2015machine}, and financial details~\cite{ngai2011application}. 
Unauthorized access to or leakage of such data can lead to significant privacy risks and adverse consequences for affected individuals.

Trained ML models can memorize and inadvertently reveal sensitive
information about their training data~\cite{song2017machine,carlini2019secret,zhang2021understanding}, making them vulnerable to several privacy attacks, such as
extraction attacks~\cite{carlini2021extracting},
property inference attacks~\cite{ganju2018property},
and membership inference attacks (MIAs)~\cite{shokri2017membership}. 

MIAs~\cite{shokri2017membership,salem2019ml,yeom2018privacy},
the focus of this paper, aim to determine whether a specific data point was part of the training data of a given model. 
Although they may not seem dangerous at first glance, they can pose serious privacy risks to individuals in specific scenarios. 
For example, knowing that a specific patient’s clinical record was used to train a model associated with a sensitive disease can reveal with high confidence that the patient suffers from this disease.

Several studies have demonstrated a strong connection between training data memorization and the
phenomenon of overfitting~\cite{yeom2018privacy,carlini2019secret,wei2024memorization}. 
Overfitting occurs when a model not only learns general patterns, but also captures sample-specific details and noise, which
leads to a noticeable difference in its behavior in training data (\textit{members}) compared to unseen data
(\textit{non-members})~\cite{shokri2017membership,yeom2018privacy,hu2022membership,wei2024memorization}. 
MIAs leverage this differential behavior~\cite{shokri2017membership,nasr2019comprehensive,song2021systematic,carlini2022membership}.

Various defenses against MIAs have been proposed and can be categorized into certified and practical defenses. 
Certified defenses provide formal privacy guarantees through differential privacy (DP)~\cite{abadi2016deep},
but often result in reduced model utility and high computational costs. 
Practical defenses, on the other hand, offer empirical privacy protection with the goal of maintaining the utility of the model~\cite{szegedy2016rethinking,nasr2018machine,jia2019memguard,song2021systematic,blanco2022critical}.
These practical defenses primarily aim to mitigate overfitting and develop
models with better generalization capabilities,
thus reducing the effectiveness of MIAs while preserving utility.
However, even models designed to generalize well can inadvertently leak information about a small portion of the training data, making them vulnerable to MIAs~\cite{long2020pragmatic,carlini2022membership}.

\textbf{Contributions}: In this paper, we address two key questions: 
\textit{Q1: What makes certain samples vulnerable to MIAs even in non-overfitted models?} and  
\textit{Q2: How can these samples be effectively protected?}

To answer these questions, we performed experiments on various data sets and models to identify factors that contribute to the vulnerability of MIA beyond overfitting.
We systematically characterize what makes samples vulnerable through visual analysis, feature-space geometry, and model explanation techniques.
We find that outliers---samples that are far from their class centroid---are particularly vulnerable. 
We then suggest and discuss potential defensive strategies to protect these vulnerable samples and thereby enhance privacy.

The remainder of this paper is organized as follows. 
Section~\ref{sec:background} provides
background on ML overfitting and differential privacy.
Section~\ref{sec:related_work} discusses related work on membership inference attacks and defenses, and factors that contribute to the success of MIAs. 
Section~\ref{sec:setup} describes the data sets, models, and experimental setup.
Section~\ref{sec:results} empirically investigates the causes of MIA beyond overfitting and discusses the results obtained. 
Section~\ref{sec:solutions} discusses potential solutions for protecting vulnerable samples.
Section~\ref{sec:conclusion} summarizes our findings and suggests future research directions.
Additional experimental details are provided in the \textbf{\href{https://github.com/najeebjebreel/mia_analysis/blob/main/mias_beyond_overfitting_supp.pdf}{supplementary materials}}.

\section{Background}
\label{sec:background}

\subsection{Machine Learning Overfitting}
\label{sec:ml}
In this paper, we focus on predictive deep neural network (DNNs) utilized as \( m \)-class classifiers, with the cross-entropy (CE) loss:
\begin{equation}
\mathcal{L}(F_\theta, z) = - \sum_{i=0}^{m-1} y_i \log(F_\theta(x)_i),
\end{equation}
where \( x \) are the input features, \( y_i \) is the one-hot encoded label vector, and \( F_\theta(x)_i \) is the predicted probability for class \( i \).

One of the potential problems of ML training is overfitting. 
Overfitting is an undesirable training outcome in which the model fits too closely to the training data but performs poorly on the test data, resulting in a high generalization error~\cite{lecun2015deep}.
Overfitting can arise from various factors, including overparameterized models, insufficient training data, high data dimensionality, or suboptimal hyperparameter selection ({\em e.g.} batch size, learning rate). 
In addition, \cite{dealcala2024comprehensive} highlight frequent data exposure during training and sharp loss functions as factors that exacerbate MIA risks.
In particular, \cite{feldman2020does} demonstrate that some degree of memorization may be essential for optimal generalization, particularly when learning from rare or unique instances.

Since best practices of ML emphasize avoiding overfitting to enhance generalization and maximize utility, our work focuses on identifying training samples that remain vulnerable to MIAs even in non-overfitted models. 

\subsection{Differential Privacy (DP)}
\label{sec:dp}
Differential privacy (DP~\cite{dwork2006differential}) ensures that the inclusion or exclusion of a single data point in a data set does not significantly affect the output of a statistical function. Formally, a mechanism \( M \) satisfies \((\epsilon, \delta)\)-DP if, for any two neighboring data sets \( D \) and \( D' \) (differing by one data point) and any subset \( S \) of outcomes:
\begin{equation}
\Pr[M(D) \in S] \leq e^{\epsilon} \Pr[M(D') \in S] + \delta,
\end{equation}
where \(\epsilon\) is the privacy budget (smaller values imply stronger privacy), and \(\delta\) is the probability of exceeding the budget.

In DNN training, DP is typically implemented via DP-SGD~\cite{abadi2016deep}, which clips per-example gradients to bound sensitivity and adds Gaussian noise to the batch gradient during training. 
However, DP-SGD introduces challenges, including complex hyperparameter tuning, increased training time, and reduced model utility~\cite{ponomareva2023dp,blanco2022critical}.

\section{Related Work}
\label{sec:related_work}

\subsection{Black-box MIA Approaches}
\label{sec:bbox_mias}
We focus on black-box MIAs since, on the one hand, according to \cite{sablayrolles2019white} they are (or can be) as good as any white-box MIAs.
There are several approaches to conducting black-box MIAs, each leveraging different aspects of the model output to distinguish between members and non-members.
Shadow model attacks~\cite{shokri2017membership} train multiple models to mimic the target model behavior, and train an ML attack model on the predictions of the shadow models to distinguish members from non-members. 
\cite{yeom2018privacy} infer a sample as a member if its loss is less than the average training loss. 
\cite{salem2019ml} threshold the confidence score of a sample to infer membership, with higher confidence indicating membership. 
\cite{song2021systematic} utilize prediction entropy, with lower entropy indicating membership. 
The likelihood ratio attack (LiRA) of \cite{carlini2022membership} applies hypothesis testing using Gaussian distributions fitted to the output of multiple models (trained with and without the target samples), achieving more reliable detection, but requiring extensive computation.

\subsection{Defenses Against Membership Inference Attacks}
\label{sec:defenses}
To mitigate membership inference attacks (MIAs), various defenses have been proposed. 
Differential privacy (DP) methods, such as DP-SGD (noise-added gradient descent)~\cite{abadi2016deep} and PATE (ensemble training with noisy voting)~\cite{papernot2018scalable}, provide formal privacy guarantees, but often reduce model utility and increase computational costs~\cite{shokri2017membership,yeom2018privacy,rahimian2020sampling,jayaraman2019evaluating}.

Anti-overfitting strategies, which maintain better utility while mitigating MIAs, include early stopping~\cite{caruana2000overfitting,song2021systematic} and regularization techniques: L2 regularization penalizes large parameters; dropout randomly deactivates units during training~\cite{srivastava2014dropout,salem2019ml}; adversarial regularization~\cite{nasr2018machine} modifies the loss function; and label smoothing replaces hard labels with soft distributions~\cite{szegedy2016rethinking}.

Output masking defenses restrict prediction details by releasing only top-k probabilities or class labels~\cite{shokri2017membership}, though top-k leakage remains a limitation. MemGuard~\cite{jia2019memguard} further perturbs confidence scores to confuse attackers. 
Knowledge distillation methods, such as DMP (low-entropy training)~\cite{shejwalkar2021membership} and SELENA (sub-model distillation)~\cite{tang2022mitigating}, transfer knowledge from teacher to student models to enhance privacy.

\subsection{Understanding MIA Vulnerabilities Beyond Overfitting}
\label{sec:understand_mia}

While overfitting is a known primary cause of MIA vulnerabilities~\cite{yeom2018privacy}, privacy leakage may also occur in non-overfitted models~\cite{long2020pragmatic,carlini2022membership}.
\cite{long2020pragmatic} identify vulnerable samples in well-generalized models as those with few neighbors in the intermediate feature space.
\cite{carlini2022membership} use shadow model training to model per-example loss distributions for members and non-members as Gaussian distributions and detect members via likelihood ratio tests.
\cite{liu2022membership} analyze loss trajectories during training to identify vulnerable samples.

Our work differs from these studies in two key ways:
(1) We provide a comprehensive visual and geometric analysis of vulnerable samples using t-SNE visualizations~\cite{maaten2008visualizing}, combined with model explanation techniques (Grad-CAM~\cite{selvaraju2017grad}) to reveal why specific samples are vulnerable. We show that models tend to focus on non-relevant features for outlier samples relative to their class centroids.
(2) Whereas prior work~\cite{long2020pragmatic,carlini2022membership,liu2022membership} primarily informs attack design, we leverage our analysis to suggest suitable defenses and propose a novel logit-reweighting method specifically targeting geometrically identified vulnerable samples.

\section{Experimental Setup}
\label{sec:setup}

\noindent \textbf{Data sets and models.} We used two benchmark data sets commonly used in the literature of MIAs, namely {\em Purchase100}~\cite{shokri2017membership} and {\em CIFAR-10}~\cite{krizhevsky2009learning}. 
For Purchase100, we used a fully connected network (FCN) as in~\cite{shejwalkar2021membership}.
For CIFAR-10, we employed two convolutional neural network architectures: DenseNet-12~\cite{huang2017densely} and ResNet-18~\cite{he2016deep}.
The utility of the model was measured through accuracy.

\noindent \textbf{Attacks and defenses.} We evaluated two black-box MIAs (loss-based~\cite{yeom2018privacy}, entropy-based~\cite{salem2019ml}) using AUC and the attacker’s advantage~\cite{yeom2018privacy}. 
MIA AUC measures the overall attack performance across all decision thresholds using the Area Under the ROC Curve. An AUC of 50\% indicates random guessing (perfect privacy), while higher values indicate more effective attacks and greater privacy leakage.
MIA attacker's advantage is defined as $2 \cdot \Pr[\text{correct guess}] - 1$~\cite{yeom2018privacy}, which is equivalent to $\max_{\tau}(\text{TPR}(\tau) - \text{FPR}(\tau))$ across all decision thresholds $\tau$. An advantage of 0\% means no benefit over random guessing, while higher percentages indicate greater privacy violations.

In addition, we identified the most vulnerable samples as true positive samples (TP) at a low false positive rate (FPR), as suggested in~\cite{carlini2022membership}; these are the samples that are the most reliably detectable by the attacker.
We considered the following defenses: early stopping~\cite{song2021systematic}, L2-regularization~\cite{shokri2017membership}, regularization and dropout (RegDrop)~\cite{blanco2022critical}, label smoothing (LS)~\cite{szegedy2016rethinking}, and DP-SGD~\cite{abadi2016deep}.

More details on system specifications, data set descriptions, models, attacks, defenses, and training settings are provided in \textbf{\href{https://github.com/najeebjebreel/mia_analysis/blob/main/mias_beyond_overfitting_supp.pdf}{supplementary materials}}.

\section{Results and Discussion}
\label{sec:results}

In this section, we first study the impact of overfitting on model utility and MIAs.
Then, we apply a set of representative defenses against MIAs (described in Section~\ref{sec:setup}) and analyze their impact on the utility of the model and MIAs.
After that, we analyze why some training samples of 
non-overfitted models are still vulnerable to MIAs.

\subsection{Impact of Overfitting}
Overfitting has an impact on the utility of the model and the effectiveness of MIAs. Let us examine this impact in depth.

\begin{figure}[!ht]
\centering
    \begin{subfigure}[b]{0.33\textwidth}
        \centering
        \includegraphics[width=\textwidth]{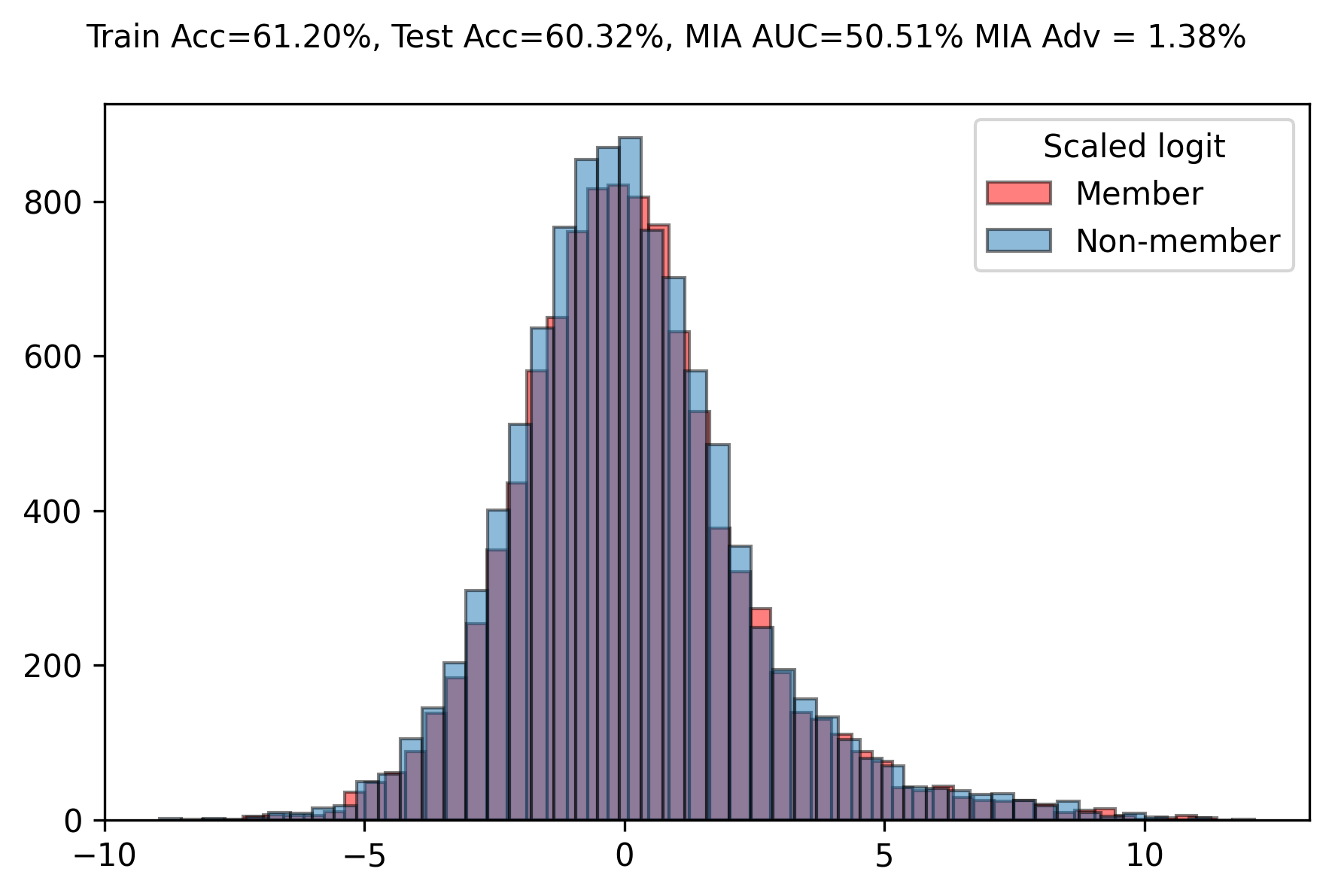}
        \caption{Epoch 1}
    \end{subfigure}%
    \begin{subfigure}[b]{0.33\textwidth}
        \centering
        \includegraphics[width=\textwidth]{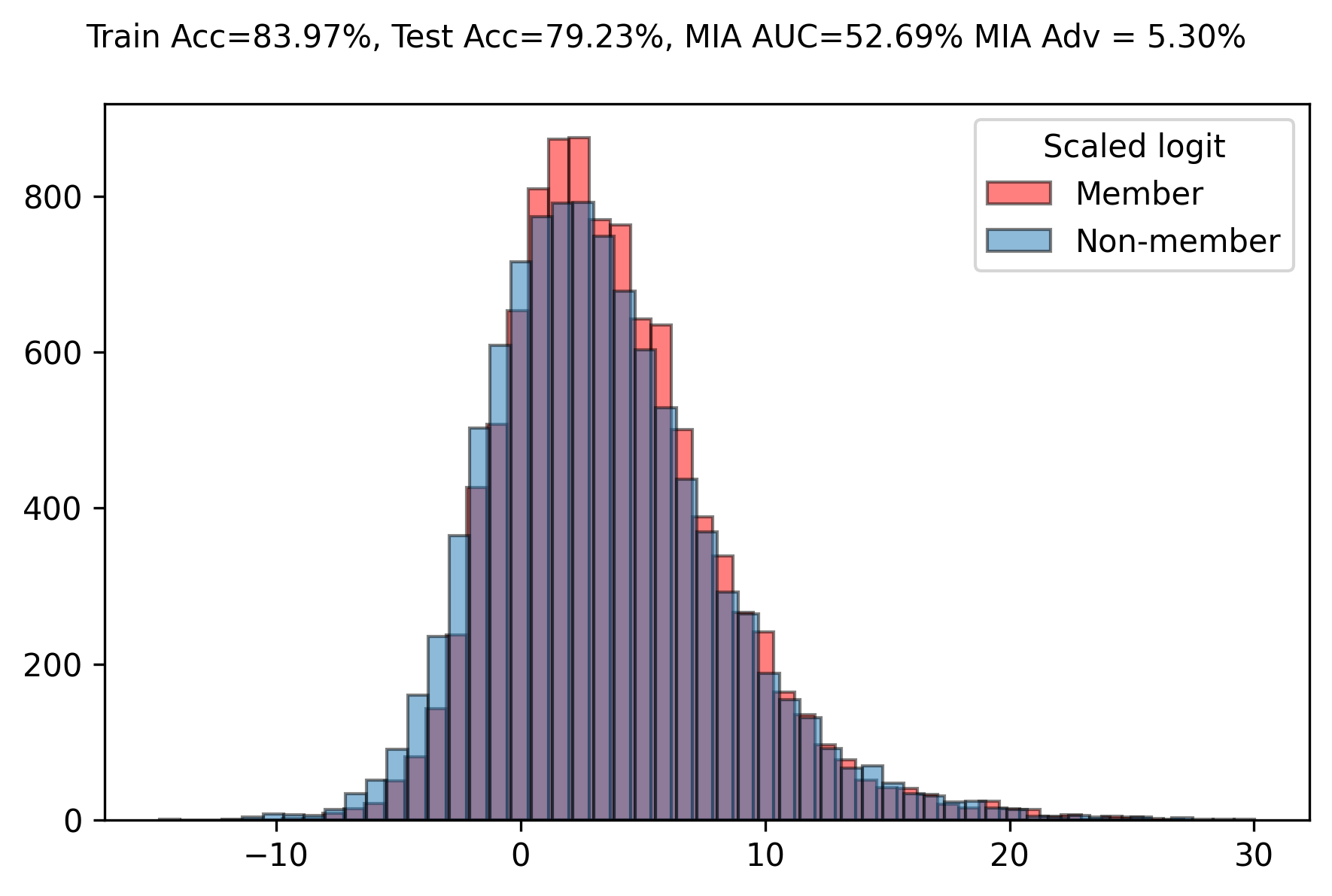}
        \caption{Epoch 5}
    \end{subfigure}%
    \begin{subfigure}[b]{0.33\textwidth}
        \centering
        \includegraphics[width=\textwidth]{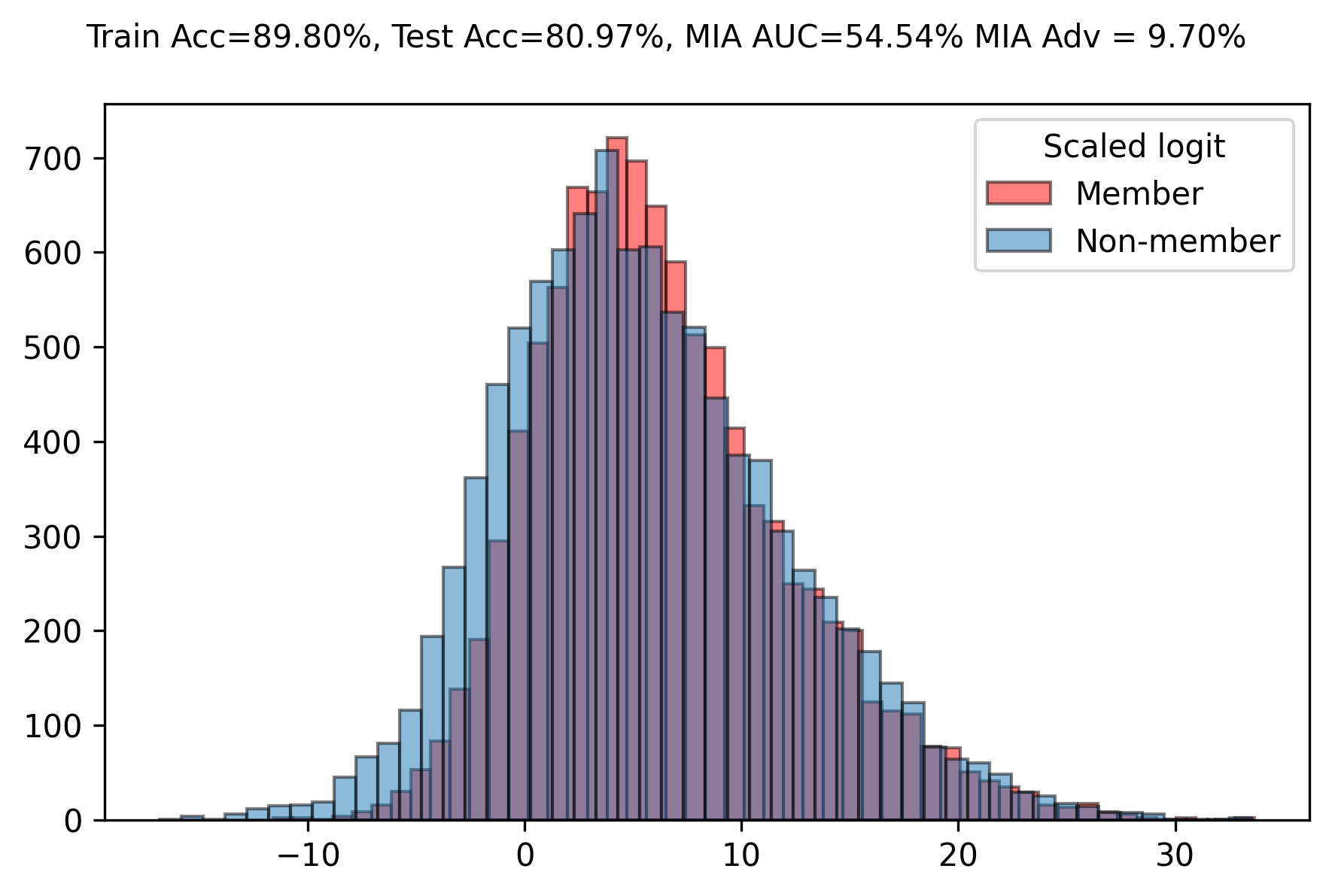}
        \caption{Epoch 10}
    \end{subfigure}
    
    \begin{subfigure}[b]{0.33\textwidth}
        \centering
        \includegraphics[width=\textwidth]{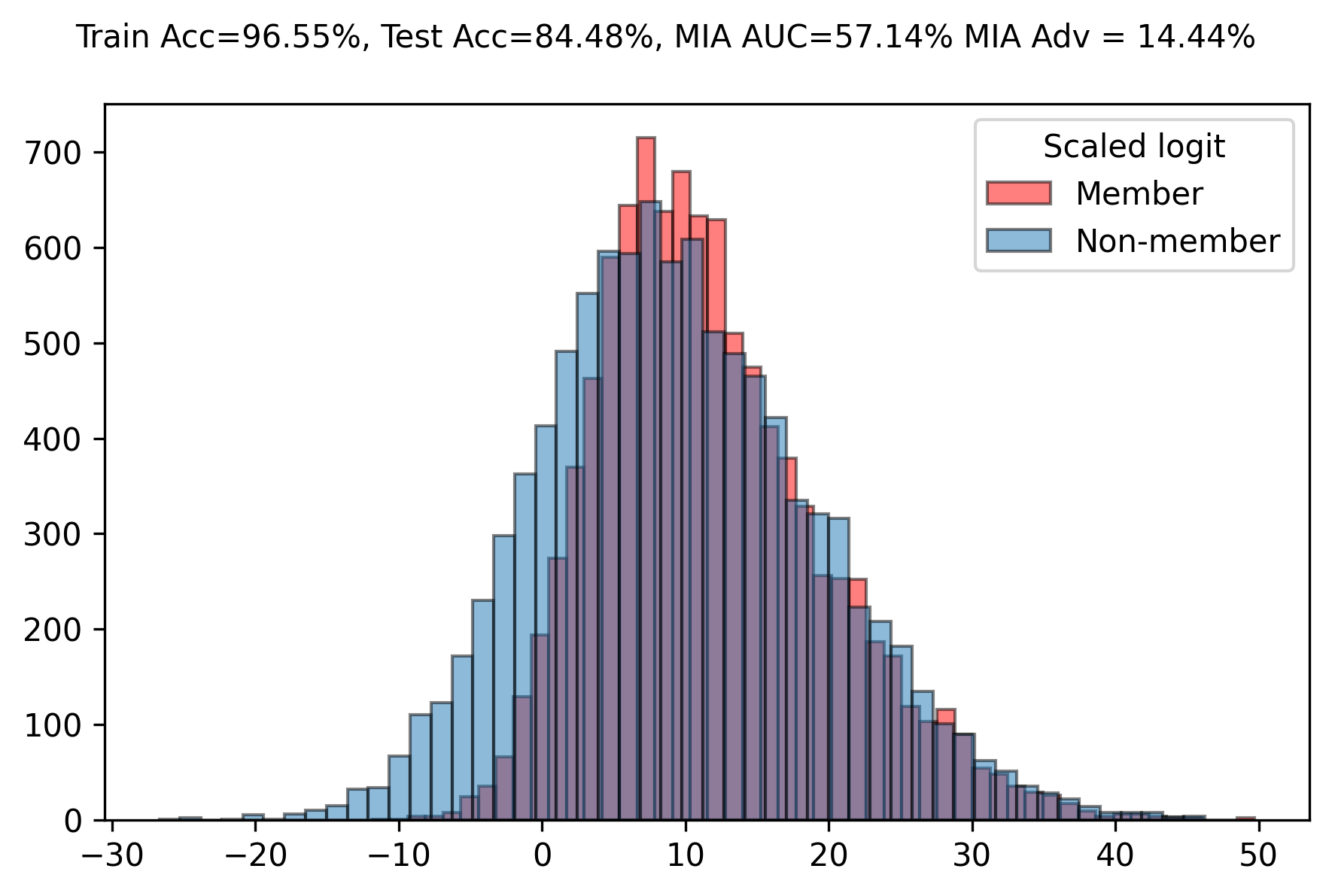}
        \caption{Epoch 20}
    \end{subfigure}%
    \begin{subfigure}[b]{0.33\textwidth}
        \centering
        \includegraphics[width=\textwidth]{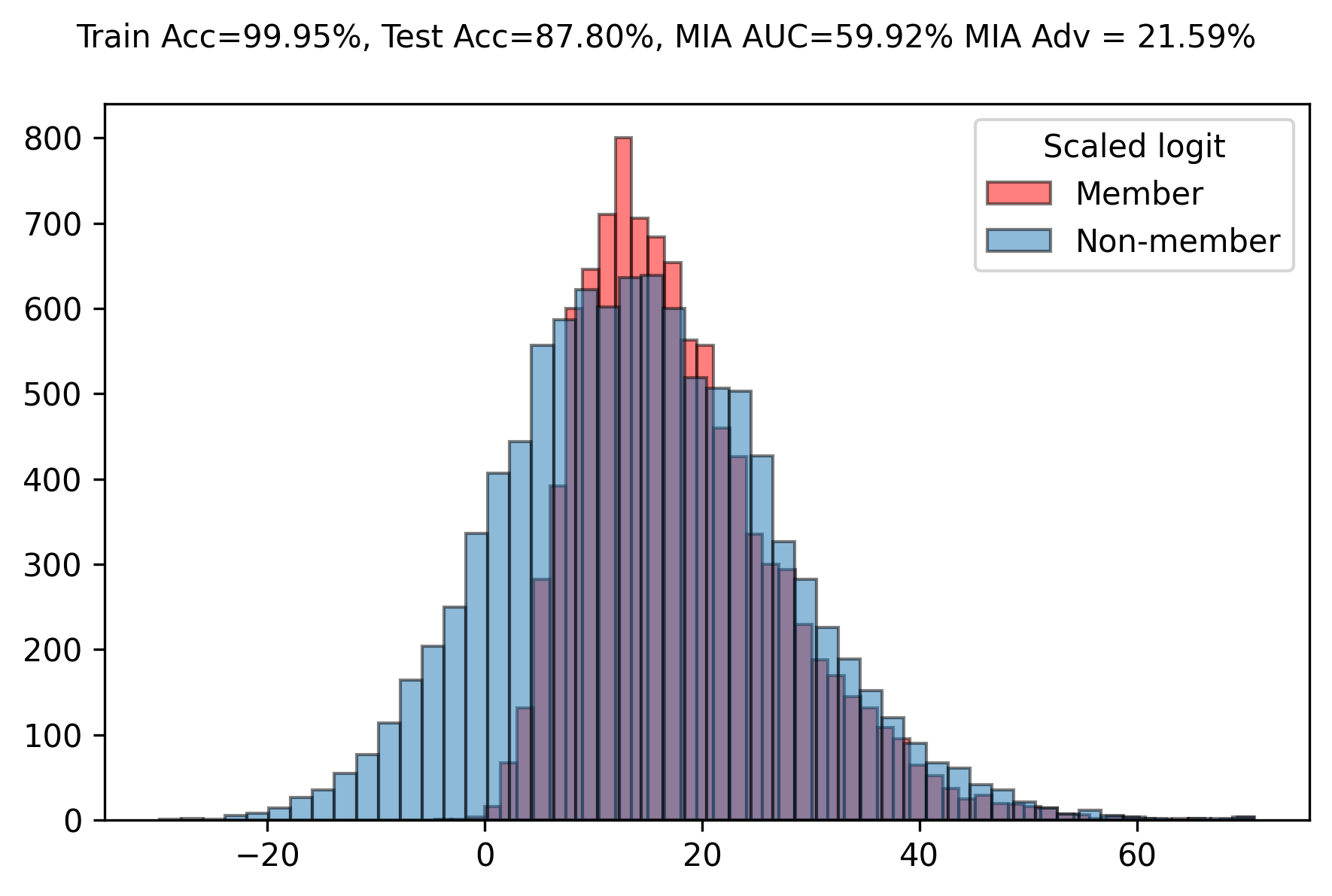}
        \caption{Epoch 40}
    \end{subfigure}%
    \begin{subfigure}[b]{0.33\textwidth}
        \centering
        \includegraphics[width=\textwidth]{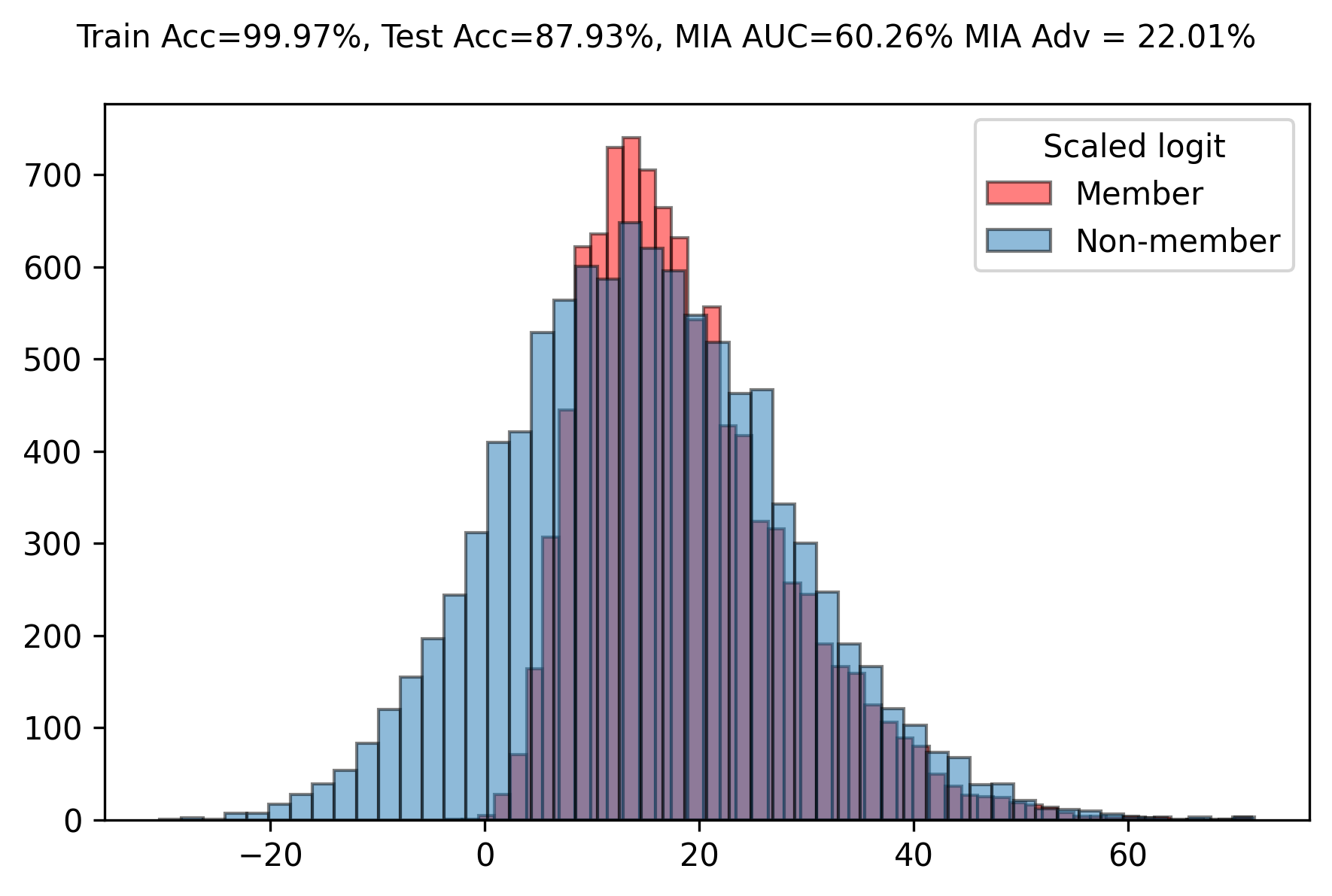}
        \caption{Epoch 60}
    \end{subfigure}

\caption{Impact of overfitting in CIFAR10-DenseNet. Distributions of scaled logits for member and non-member data points, accuracy metrics, and MIA metrics for several epochs.}
\label{fig:overfitting_impact_densenet}
\end{figure}

\noindent \textbf{Separation between members and non-members.}
Figure~\ref{fig:overfitting_impact_densenet} illustrates the histograms of the distributions of scaled logits~\cite{carlini2022membership} for member and non-member data points across different epochs during the training of the CIFAR10-DenseNet-12. 
The figure also displays metrics related to model utility, namely training accuracy (Train Acc) and test accuracy (Test Acc), as well as metrics related to membership inference attacks (MIA), specifically MIA AUC and MIA attacker advantage (MIA Adv). 
These metrics indicate that as the model trains and begins to overfit, the gap between training and test accuracy increases, and the separation between member and non-member data points becomes more pronounced, thereby increasing the model's vulnerability to MIAs.

\noindent \textbf{Overfitting and model complexity.} Table~\ref{tab:model_complexity} compares accuracy and MIA metrics for two models whose number of parameters are significantly different.  
It can be seen that in the larger model the gap between training and test performance is greater, which makes MIAs more effective. This is a sign of overfitting by the larger model.

\begin{table}[!ht]
\centering
\caption{Impact of model complexity}
\label{tab:model_complexity}
\resizebox{\textwidth}{!}{%
\begin{tabular}{lccccc} 
\toprule
Method   & \# params                  & Train Acc & Test Acc & MIA AUC & MIA Adv.  \\ 
\hline
DenseNet & $\sim$770,000  & 99.97     & 87.91    & 60.27   & 22.07     \\
ResNet   & $\sim$11,170,000 & 99.26     & 82.79    & 64.45   & 28.31     \\
\bottomrule
\end{tabular}}%
\end{table}

\subsection{Effectiveness of Defenses Against MIAs} 
This section evaluates several defense mechanisms (described in
Section~\ref{sec:defenses}) designed to mitigate the vulnerability of DNN models to membership inference attacks. 
These defenses are tested on two benchmarks: Purchase100-FCN and CIFAR10-DenseNet-12. 
The performance of these defenses is evaluated in terms of utility, runtime, and resistance to MIAs.

An ideal defense should maintain or exceed the model's original accuracy
(due to improved generalization) with a similar or lower runtime. In terms of privacy protection, an optimal defense should render MIAs as ineffective as random guessing, achieving an AUC of 50\% and a zero advantage in predicting membership status.

\begin{table*}[!ht]
\centering
\caption{Performance of defenses with Purchase100-FCN. Best figures are boldfaced, second-best are underlined.}
\label{tab:purchase_mlp}
\resizebox{\textwidth}{!}{%
\begin{tabular}{lccccc} 
\toprule
Method                      & Train Acc (\%)      & Test Acc  (\%)     & Runtime (s)  & MIA AUC (\%)        & MIA Adv. (\%)        \\ 
\hline
Original                        & \uline{97.76}  & 87.54          & \uline{1201} & 57.27          & 13.86                   \\
Early stopping                         & 96.88          & \textbf{89.58} & \textbf{200} & 55.07          & 10.40                 \\
Regularization($\lambda$=5e-4)                & 94.91          & \uline{89.34}  & 1209         & 53.25          & 7.26           \\
Regularization($\lambda$=1e-3)                & 92.63          & 88.37          & 1205         & 52.22          & 4.87             \\
Regularization($\lambda$=5e-3)                & 77.76          & 76.16          & 1207         & \uline{50.92}          & \uline{1.73}             \\
RegDrop($\lambda$=5e-4,dr=0.25) & 90.02          & 87.14          & 1489        & 51.87         & 3.70                     \\
RegDrop($\lambda$=5e-4,dr=0.50)  & 86.52          & 84.45          & 1320       & 51.44         & 2.46            \\
Label smoothing                         & \textbf{99.15} & 88.52          & 1699                & 59.43          & 16.43                   \\
DP($\epsilon=2.38$)                & 61.71          & 61.21          & 3507                     & \textbf{50.36} & \textbf{0.70}            \\
\bottomrule
\end{tabular}}%
\end{table*}

Table~\ref{tab:purchase_mlp} shows the performance of defenses with the Purchase100-FCN benchmark.
The original model achieved a high training accuracy of 97.76\% and a test accuracy of 87.54\%. 
However, it showed vulnerability to MIAs with MIA AUC 57.27\% and MIA advantage 13.86\%. 

Early stopping achieved the best test accuracy to 89.58\% and the shortest runtime (200 seconds).
It also slightly decreased the MIA AUC and advantage to 55.07\% and 10.40\%, respectively. 
This is because early stopping in this benchmark managed to stop the model training process before seriously overfitting the training data.

Regularization with different \( \lambda \) values showed a trend of degrading accuracy and improving privacy as the regularization strength increased. 
Regularization with $\lambda=5e-4$ improved test accuracy to 89.34\%, reduced MIA AUC to 53.25\%, and MIA advantage to 7.26\%. 
Increasing $\lambda$ to $1e-3$ further reduced the MIA AUC and advantage to 52.22\% and 4.87\%, respectively, with a slight drop in test accuracy to 88.37\%. 
The highest regularization ($\lambda=5e-3$) significantly reduced both training and test accuracy (77.76\% and 76.16\%), but achieved the lowest MIA AUC (50.92\%) and MIA advantage (1.73\%). 
This indicates a strong trade-off between model performance and privacy, where higher regularization reduces overfitting and enhances privacy at the cost of accuracy.
Regularization also took a runtime similar to that of the original training.
These results suggest that regularization with $\lambda = 1e-3$ struck the best balance between utility, runtime and privacy for this benchmark.

RegDrop with $\lambda=5e-4$ and dropout rates 0.25 and 0.50 slightly degraded utility, but significantly reduced MIA effectiveness. For dropout rate 0.25 we obtained test accuracy 87.14\%, MIA AUC 51.87\%, and MIA advantage 3.70\%.  
Increasing the dropout rate to 0.50 reduced test accuracy to 84.45\% but further lowered the MIA AUC to 51.44\% and the MIA advantage to 2.46\%. 
These results suggest that RegDrop offered the best balance between utility and privacy for this benchmark.

Label smoothing achieved relatively high test accuracy (88.52\%). 
However, it increased the model susceptibility to MIAs, as reflected by the MIA AUC of 59.43\% and advantage of 16.43\%. 
This result indicates that, while label smoothing increases training accuracy, it may cause the model to leave a distinguishable pattern in predictions of training samples, thus exacerbating the vulnerability to MIAs.

Differential privacy with $\epsilon=2.38$ drastically reduced the MIA AUC to 50.36\% and MIA advantage to 0.70\%, offering the strongest defense against MIAs. 
However, this came at the expense of model utility, because training and test accuracy dropped to 61.71\% and 61.21\%, respectively. 
The significant accuracy reduction highlights the trade-off of DP between strong privacy guarantees and model utility. 
The runtime (3507 seconds) was also the highest, indicating a substantial computational cost to reach convergence when training under DP.

In summary, we can see diverse trade-offs between model utility, computational cost, and privacy among defenses. 
Early stopping provided the best balance between utility and runtime.
However, it only slightly mitigated MIAs. 
{\em Moderate regularization showed the best utility-runtime-privacy trade-off among all defenses for this benchmark.}  
RegDrop, particularly at a low dropout rate, offered the best balance between utility and privacy. It achieved privacy protection close to that of DP with much better utility and runtime. 
Although differential privacy provided the strongest privacy protection, this came at the cost of significant accuracy loss and increased runtime. 
An interesting note is that regularization with $\lambda = 5e-3$ achieved an effectiveness against MIAs close to that of DP but with much better accuracy and runtime.
Label smoothing, despite its high training accuracy, increased MIA vulnerability, suggesting that its application requires careful tuning.

\begin{table*}[!ht]
\centering
\caption{Performance of defenses with CIFAR10-DenseNet-12. Best figures are boldfaced, second-best are underlined.}
\label{tab:cifar10_densenet}
\resizebox{\textwidth}{!}{%
\begin{tabular}{lccccc} 
\toprule
Method                      & Train Acc (\%)      & Test Acc (\%)       & Runtime (s)   & MIA AUC (\%)        & MIA Adv. (\%)       \\ 
\hline
Original                        & \uline{99.97}  & 87.91          & \uline{3558}  & 60.27          & 22.07          \\
Early stopping                           & \uline{99.97}  & 87.93          & \textbf{2024} & 60.21          & 21.96          \\
Regularization($\lambda$=5e-4)                & \textbf{99.99} & \uline{91.46}  & 3573          & 57.11          & 19.13          \\
Regularization($\lambda$=1e-3)                & 99.95          & 89.61          & 3564          & 58.07          & 20.52          \\
Regularization($\lambda$=5e-3)               & 60.35          & 59.71          & 3574          & \uline{50.06}  & \uline{0.82}   \\
RegDrop($\lambda$=5e-4,dr=0.25) & 99.85          & \textbf{91.78} & 3616          & 56.00          & 15.44          \\
RegDrop($\lambda$=5e-4,dr=0.50)  & 91.97          & 84.89          & 3610          & 53.61          & 7.52           \\
Label smoothing                           & \textbf{99.99} & 86.47          & 3539          & 67.33          & 37.04          \\
DP($\epsilon=4.95$)                & 59.51          & 59.55          & 7738          & \textbf{50.00} & \textbf{0.53}  \\
\bottomrule
\end{tabular}}
\end{table*}

Table~\ref{tab:cifar10_densenet} reports the same defense analysis for the CIFAR10-DenseNet-12 benchmark.
The results show that the original model achieved an extremely high training accuracy (99.97\%) and a test accuracy 87.91\%. 
However, the high MIA AUC (60.27\%) and advantage (22.07\%) indicate overfitting, making the model vulnerable to MIAs.

Early stopping maintained a high training accuracy (99.97\%) and slightly improved the test accuracy to 87.93\%. It also reduced the runtime significantly to 2024 seconds.
However, resistance to MIAs was not improved.

Regularization with $\lambda=5e-4$ achieved the second highest test accuracy (91.46\%). 
It also slightly improved privacy with MIA AUC 57.11\% and advantage 19.13\%. 
Increasing regularization to $\lambda=1e-3$ slightly improved test accuracy (89.61\%) and  privacy metrics (MIA AUC 58.07\% and advantage 20.52\%). 
At the highest regularization strength ($\lambda=5e-3$), there was a drastic drop in both training and test accuracy (60.35\% and 59.71\%, respectively), but this setting achieved the lowest MIA AUC (50.06\%) and advantage (0.82\%), indicating strong privacy protection at the cost of performance. 

RegDrop with $\lambda=5e-4$ and a dropout rate of 0.25 achieved the best balance, with the highest test accuracy (91.78\%) and improved privacy metrics (MIA AUC 56.00\% and advantage 15.44\%). 
Increasing the dropout rate to 0.50 reduced test accuracy to 84.89\% but further enhanced privacy (MIA AUC 53.61\% and advantage 7.52\%). 
This demonstrates RegDrop's effectiveness in mitigating overfitting and enhancing privacy while maintaining reasonable accuracy.

Label smoothing achieved the highest training accuracy (99.99\%) but offered a lower test accuracy (86.47\%) compared to the baseline original model. 
This method actually increased the model's vulnerability to MIAs, with the highest MIA AUC (67.33\%) and MIA advantage (37.04\%). 
This suggests that, whereas label smoothing can improve training performance, it may render the model more susceptible to privacy attacks.

Differential privacy with $\epsilon=4.95$ provided the strongest defense against MIAs, achieving the lowest MIA AUC (50.00\%) and advantage (0.53\%), almost similar to random guessing. 
However, this came with significant reductions in both training and test accuracies (59.51\% and 59.55\%, respectively) and a substantial runtime (7738 seconds).

In summary, regularization and its combination with dropout achieved the best utility-privacy balance. 
Early stopping offered computational efficiency, but weak privacy protection. 
Differential privacy provided strong formal guarantees, but degraded performance and increased computational costs. 
Label smoothing improved training accuracy but increased MIA vulnerability. 
These findings suggest that selecting appropriate regularization and dropout parameters is the most effective approach to balancing utility, runtime, and privacy when training DNN models, as also observed by~\cite{blanco2022critical}.

Despite its strong generalization capabilities, the best-performing CIFAR10-DenseNet-12 model (RegDrop with $\lambda=5e-4$ and a dropout rate 0.25) still exhibited an MIA AUC 56.00\% and an MIA advantage 15.44\%, which are both above the level expected from random guessing. This raises the crucial question addressed in
this paper: \textbf{Why do membership inference attacks perform better than random guessing on models exhibiting good generalization, and what characteristics define the training samples that remain vulnerable to MIAs?}  
To answer this question, the following section provides a thorough examination of the training samples that continue to be susceptible to MIAs even after successfully mitigating overfitting in the CIFAR10-DenseNet-12 model (RegDrop with $\lambda=5e-4$ and a dropout rate of 0.25).

\subsection{Vulnerable Samples Beyond Overfitting}

We focus on the most vulnerable training samples, selecting true positives (TP) with a 1\% false positive rate (FPR) following~\cite{carlini2022membership}. 
We directly used the loss values of the train and test samples from the target model.  

\begin{figure}[t!]
\centering
    \begin{subfigure}[b]{0.5\textwidth}
        \centering
        \includegraphics[width=\textwidth]{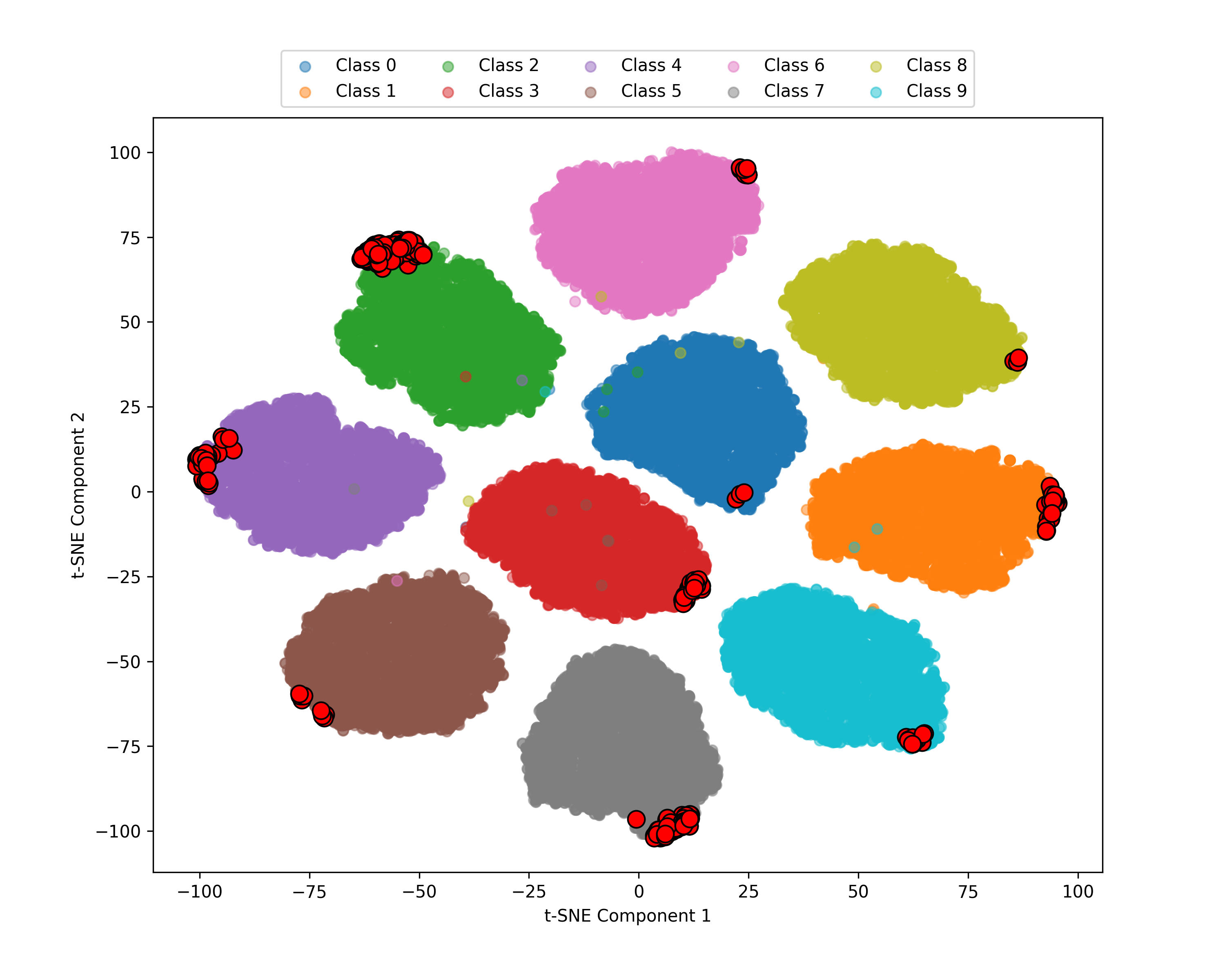}
        \caption{TP @ 1\% FPR}
        \label{fig:tsne_tp_1perfpr}
    \end{subfigure}%
    \begin{subfigure}[b]{0.5\textwidth}
        \centering
        \includegraphics[width=\textwidth]{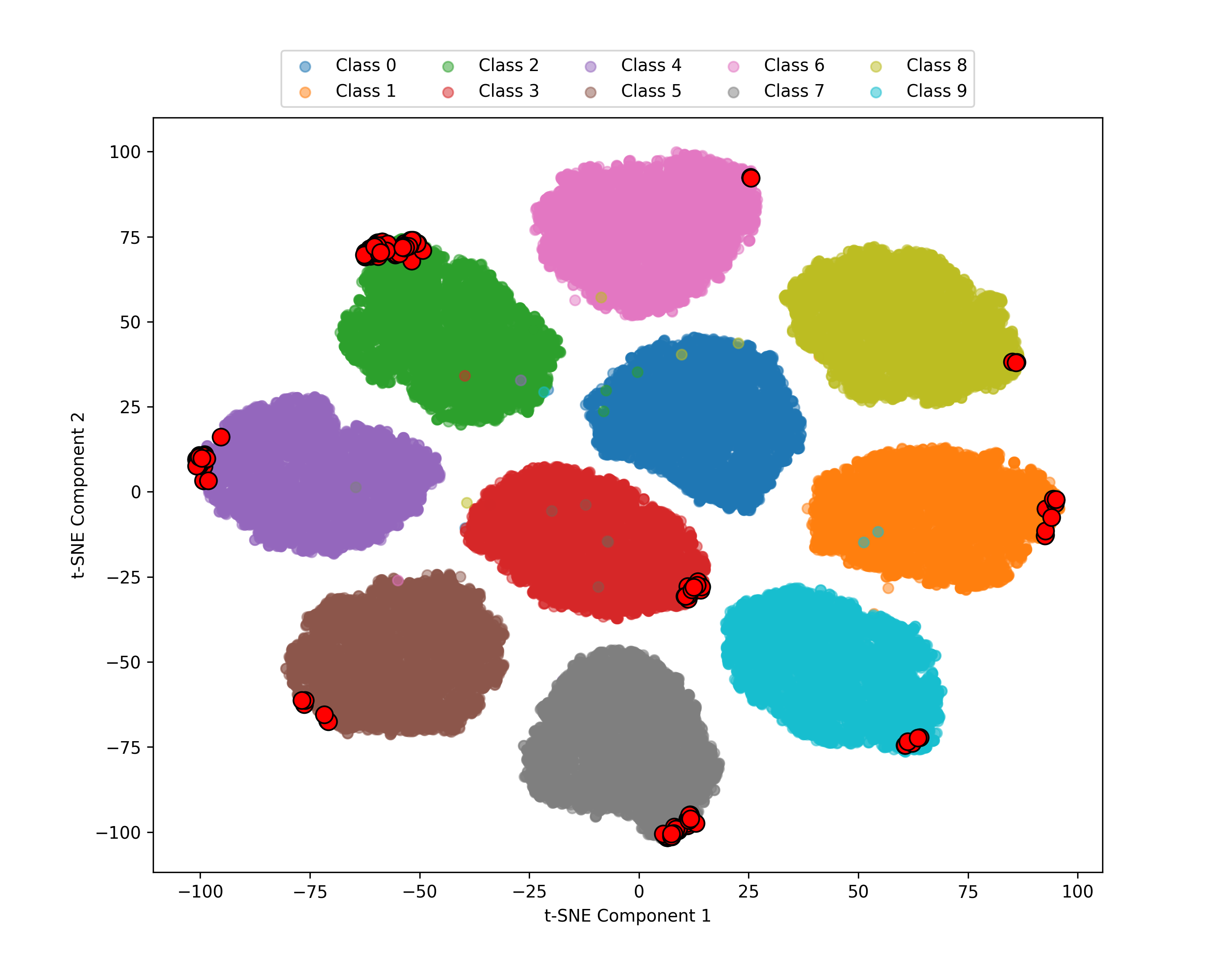}
        \caption{TP @ 0.5\% FPR}
        \label{fig:tsne_tp_0.5perfpr}
    \end{subfigure}
\caption{t-SNE visualization of vulnerable samples (circled red)  w.r.t their class samples}
\label{fig:tsne_vulnerable}
\end{figure}

The t-SNE visualization of the latent features of these samples in Figure~\ref{fig:tsne_tp_1perfpr} shows that these vulnerable samples are located primarily on the borders of their respective class clusters. This suggests that these samples differ significantly from the majority, likely being hard-to-classify, noisy, or outliers. 
Such characteristics may cause the model to memorize these samples based on specific details rather than relevant class patterns, leading to overconfidence in predictions and increased vulnerability to MIAs.
True positives (TP) with a false positive rate of 0.5\% (FPR) are also shown in Figure~\ref{fig:tsne_tp_0.5perfpr}.

\begin{figure*}[t!]
\centering
    \begin{subfigure}[b]{\textwidth}
        \centering
        \includegraphics[width=\textwidth]{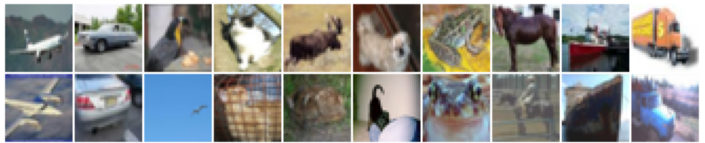}
        \caption{Visualization of samples close to the class centroid (first row) and vulnerable samples (second row). Samples were taken from the CIFAR10 training set.}
        \label{fig:outlier_samples}
    \end{subfigure}%
    
    \begin{subfigure}[b]{\textwidth}
        \centering
        \includegraphics[width=\textwidth]{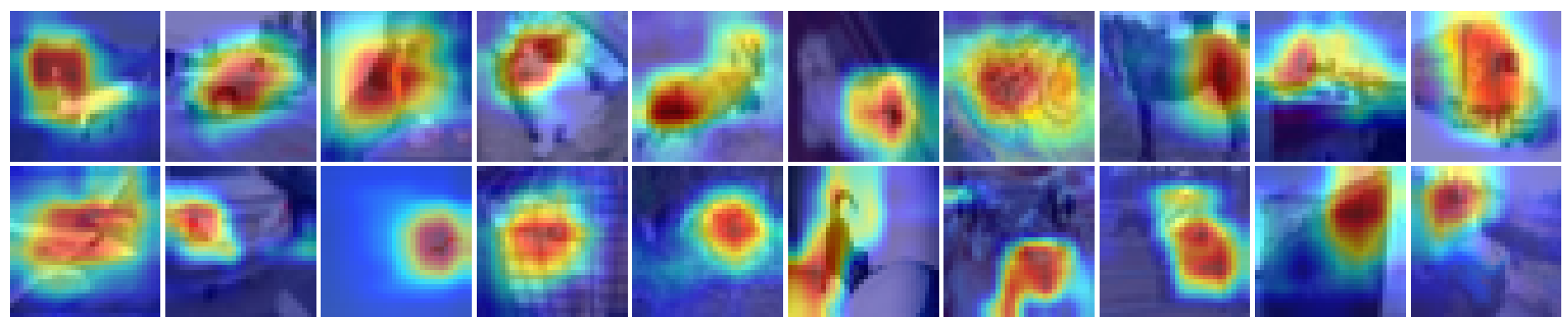}
        \caption{Grad-CAM explanations of samples close to the class centroid (first row) and vulnerable samples (second row).}
        \label{fig:outlier_gradcam}
    \end{subfigure}

\caption{Visualization of protected and vulnerable samples and their explanations}
\label{fig:outlier_samples_gradcam}
\end{figure*}

To further investigate the nature of these boundary samples that remain vulnerable to MIAs, we analyze their characteristics compared to typical class samples. 
For CIFAR10-DenseNet-12,
Figure~\ref{fig:outlier_samples_gradcam} provides visualizations and explanations of samples close to the class centroid and vulnerable samples. Explanations are based on
the Grad-CAM method (gradient-weighted class activation mapping, \cite{selvaraju2017grad}), which highlights the pixels responsible
for decisions. Specifically:
\begin{itemize}
\item Figure~\ref{fig:outlier_samples} shows the inlier images close to their centroids (first row) and the images most vulnerable to MIAs from each class (second row) .
We can see that inlier samples are clear and easy to classify while the most vulnerable samples are noisy ({\em e.g.}, the cat hidden by the red net), unclear ({\em e.g.}, the tiny bird in the blue sky and the man riding the horse), or hard to classify ({\em e.g.}, the black cat and the big face frog).
\item Figure~\ref{fig:outlier_gradcam} gives the Grad-CAM explanations of the classification decisions for the images in Figure~\ref{fig:outlier_samples}.
In the case of the inlier images, the relevant pixels corresponding to the class's general patterns were identified. 
For the vulnerable examples, non-relevant pixels were generally identified. In most cases, these identified pixels were related to noise details ({\em e.g.}, the red net obscuring the cat) or sample-specific details ({\em e.g.}, the rear traffic light of the car and the people riding the truck).
\end{itemize}

These observations indicate that noisy or unclear samples may inherently resist MIAs because they do not facilitate the clear identification of individual data points. 
In contrast, clear samples with unique or untypical features —--those that are difficult to classify--— are particularly vulnerable to MIAs.
Even in a model with good generalization capability, overfitting to these unique aspects can lead to memorization, which attackers can exploit.

\section{Potential Solutions}
\label{sec:solutions}

This section explores potential defenses to mitigate the memorization of vulnerable samples in DNNs, depending on whether these samples are identified beforehand. 
Most defenses are based on established techniques, but we also introduce a novel logit-reweighting method and provide practical guidelines to protect identified vulnerable samples.

\subsection{Protection Before Identifying Vulnerable Samples}

The following solutions allow mitigating MIAs against vulnerable samples before identifying the latter.

    \noindent \textbf{Appropriate regularization and dropout.} The results in Section~\ref{sec:results} have shown the effectiveness of regularization and dropout techniques in mitigating MIA while maintaining the utility of the model. A careful choice of the regularization penalty factor $\lambda$ and the dropout ratio could further enhance the protection of these samples.
    
    \noindent \textbf{Data augmentation.} Data augmentation techniques~\cite{shorten2019survey} can be particularly effective in reducing the memorization of vulnerable samples.  By generating new training samples through transformations such as rotations, translations, scaling, and mixup~\cite{zhang2017mixup}, the model is exposed to a broader variety of data points. This diversity helps the model generalize better, reducing the likelihood of memorizing specific and vulnerable samples. 
    
     \noindent \textbf{Curriculum learning.} Curriculum learning involves gradually increasing the complexity of training data~\cite{wang2021survey}. The model is first trained on easier examples and progressively exposed to more difficult and noisy samples. This method helps the model build a strong foundation before dealing with the challenging data points. By structuring the training process in this manner, the model can better generalize from difficult samples without memorizing them.
    
     \noindent \textbf{Ensemble learning.} Ensemble learning methods combine the predictions of multiple models to improve overall performance and robustness~\cite{dong2020survey}. Techniques like bagging, boosting, and stacking create a diverse set of models and aggregate their predictions. Ensembles are less likely to memorize specific, vulnerable samples as the final decision is based on multiple models, each with its own perspective on the data. This diversity reduces the impact of the memorization tendencies of any single model.

\subsection{Protection After Identifying Vulnerable Samples}
Once vulnerable samples are identified, protection becomes easier. Potential solutions include:\\
     
     \noindent \textbf{Retraining after excluding vulnerable samples.} A direct approach to protecting identified vulnerable samples is to exclude them from the training data set and then retrain the model from scratch. Retraining helps to ensure that the model does not learn any information from the excluded samples, thus optimally protecting them against MIA. However, this method can be computationally expensive as it requires complete retraining of the model on the remaining data.

     \noindent \textbf{Machine unlearning.} Machine unlearning refers to the process of efficiently forgetting specific data points from a pre-trained ML model as if they had never been part of the training set~\cite{bourtoule2021machine,ginart2019making}. Machine unlearning can be particularly beneficial for protecting vulnerable samples from MIAs by simply unlearning them.

     \noindent \textbf{Latent feature or logit generalization.} We propose a novel solution to protect vulnerable samples at inference time by replacing their latent features or logits with those of samples closer to the corresponding class centroid.  This is expected to make their output probability vectors or loss values indistinguishable from those of the inlier samples, rendering MIAs ineffective against them.
    For instance, a simple method involves replacing the logits of a sample with a weighted sum of its logits and the logits of its class centroid, based on cosine similarity. 
    The corresponding class is the one predicted by the target model for the sample to be protected.
    Samples farther from the centroid receive higher weight adjustments. The results of this approach for CIFAR10-DenseNet-12 are shown in Table~\ref{tab:ours_cifar10_densenet}.
    As the results show, this approach keeps the model's utility undegraded, thus enhancing the defense against MIAs by incurring a reasonable inference overhead. 
    It can be seen that such a defense at inference time can also complement the performance of the anti-overfitting methods at training time ({\em e.g.}, regularization and/or dropout).
    Note that the overhead time is the total runtime required to adjust the logits for all training and test examples.
    \begin{table*}[t!]
    \centering
    \caption{Performance of the simple logit-reweighting defense with CIFAR10-DenseNet-12}
    \label{tab:ours_cifar10_densenet}
    \resizebox{\textwidth}{!}{%
    \begin{tabular}{llccccc} 
    \toprule
    \multicolumn{2}{c}{Method} & Train Acc (\%) & Test Acc (\%) & Inference Overhead (s) & MIA AUC (\%) & MIA Adv. (\%) \\ 
    \hline
    \multirow{2}{*}{Original} & Before & 99.97 & 87.91 & 0 & 60.27 & 22.07 \\
    & After & 99.97 & 87.91 & 0.462 & 55.94 & 11.89 \\ 
    \hline
    \multirow{2}{*}{RegDrop ($\lambda$=5e-4, dr=0.25)} & Before & 99.85 & 91.78 & 0 & 56.00 & 15.44 \\
    & After & 99.85 & 91.78 & 0.467 & 53.76 & 7.73 \\
    \bottomrule
    \end{tabular}}%
    \end{table*}

\section{Conclusions and Future Work}
\label{sec:conclusion}

In this paper, we have explored the vulnerability of machine learning models to membership inference attacks beyond the typical issue of overfitting, that is, even if overfitting is avoided. 
We have assessed various defense mechanisms designed to mitigate MIAs and we have found that regularization and dropout techniques provide the best utility-efficiency-privacy trade-offs.
Our investigation has revealed that even non-overfitted models with good generalization capabilities can nonetheless expose information about specific training samples, making them vulnerable to MIAs.
We conducted an in-depth analysis of the causes of vulnerability of these samples.
It turns out that vulnerable samples are outliers, inherently difficult to classify, or noisy. 
Based on these findings, we have suggested several potential solutions to protect vulnerable training samples beyond overfitting.

\textbf{Limitations:} While our findings offer valuable insights, our study has limitations that should be acknowledged. 
We focus on one tabular dataset (Purchase100) and one image dataset (CIFAR-10), which may not generalize to other domains such as text or audio. 
Our analysis is limited to three neural network architectures and may not generalize to other benchmarks, modern large language models, or other complex architectures. 
Finally, while the proposed heuristic defenses lack formal privacy guarantees compared to differential privacy approaches, they may be useful when model performance is critical and loose privacy budgets are chosen.

For future work, we will: (i) explore MIAs beyond overfitting across diverse data sets and models, 
(ii) develop dynamic defenses during training and inference to protect vulnerable samples while preserving utility, 
(iii) optimize regularization and dropout for privacy-utility trade-offs, 
(iv) assess whether excluding vulnerable samples before retraining mitigates new risks,
and (v) incorporate additional evaluation metrics, such as TPR@LowFPR, to better assess attack effectiveness.

\begin{credits}
\subsubsection{\ackname} 
This work was partly funded by 
the Centre International de Math\'ematiques et d'Informatique de Toulouse
(CIMI), the
Government of Catalonia (ICREA Acad\`emia Prize to J. Domingo-Ferrer), 
MCIN/AEI/ 10.13039/501100011033 and ``ERDF A way of making Europe'' under grant PID2021-123637NB-I00 ``CURLING'', and INCIBE and European Union NextGenerationEU/PRTR (project ``HERMES'' and INCIBE-URV Cybersecurity Chair).

\end{credits}

\bibliographystyle{splncs04}
\bibliography{my_bib}
\end{document}